\documentclass{moriond}

\usepackage{wrapfig}  

\def\be{\begin{equation}}
\def\ee{\end{equation}}
\def\bea{\begin{eqnarray}}
\def\eea{\end{eqnarray}}

\newcommand{\Photo}{}
\newcommand{\rmd}{\mathrm{d}}

\begin{document}
\vspace*{4cm}
\title{Cluster Counts Tension : a Sign of Primordial Non-Gaussianity ?}

\author{ Z. SAKR$^{1,2}$, D. HOUGUENAGUE$^{3}$, A. BLANCHARD$^{1}$}

\address{$^{1}$IRAP, Universit\'e de Toulouse, CNRS, CNES, UPS, Toulouse, France \\ $^{2}$Universit\'e St Joseph; UR EGFEM, Faculty of Sciences, Beirut, Lebanon\\$^{3}$Université d'Aix-Marseille - Facult\'e des Sciences - Marseille, France}

\maketitle\abstracts{
Evolution and abundance of the large-scale structures we observe today, such as clusters of galaxies, is sensitive to the statistical properties of dark matter primordial density fluctuations, which is assumed to follow a Gaussian probability distribution function. Within this assumption, a significant disagreement have been found between clusters counts
made by Planck and their prediction when calibrated by CMB angular power spectrum. The purpose of this work is to relax the Gaussianty assumption and test if Non-Gaussianity in dark matter primordial density fluctuations, could alleviate the tension.} 

\section{Introduction} 

$\Lambda$CDM model has proved successful in describing to a high precision most of nowadays cosmological observations \cite{Ko08}. Within its framework, galaxy clusters counts are a powerful tool to constrain
cosmological parameters \cite{o92} \cite{all11}, more specifically  the matter density $\Omega_m$ and the current amplitude of matter fluctuations, characterized by the $\sigma_8$ parameter, both being  the main ingredients entering the linear growth of structure. However, the standard $\Lambda$CDM model shows a significant disagreement between  $\sigma_8$ from clusters counts by Planck Mission and that from CMB  angular power spectrum \cite{PlkSZ16}. Beside an improper calibration of the cluster counts, with the later being calculated from a halo mass distribution function 
of the linear power spectrum, the discrepancy could be solved through extensions to $\Lambda$CDM model like massive neutrinos or modified gravity theory altering the power spectrum. These alternatives were investigated using X-ray clusters \cite{sib18}, however, within Gaussian initial fluctuations considerations, while some authors \cite{ouk97} pointed out that the shape of the probability distribution
of initial density fluctuation present in P\&S formula through the power spectrum can be replaced by  an appropriate distribution function so that to fit the local properties of clusters. This calls for considering primordial Non-Gaussianity as a potential solution for fixing the discrepancy.
Yet, few groups have tried in the past to investigate the effect of primordial Non-Gaussianity on halo mass function and cluster counts \cite{Chiu97,Tri12}. 
This motivated us in this work, to investigate if primordial Non-Gaussianity could help fix the discrepancy on $\sigma_8$.

\section{Methods }

Constraints on the cosmological parameters subject of discrepancy from cluster counts follows  from : 
\begin{equation}
\label{Cluster_abundance}
\frac{dN}{dz}\left(z,M>M_{lim}\right)=O(M) f_{sky}\frac{dV}{dz}\left(z\right)\int_{M_{lim}}^{\infty}dM\frac{dn}{dM},
\end{equation}
where $f_{sky}$ is the fraction of sky and $\frac{dV}{dz}$ the comoving volume being observed, $dn/dM$ is the mass function of the power spectrum calculated for a set of cosmological parameters and $O(M)$ a relation between cluster's mass and observable relation.\\
The mass function of cosmological structures from initially Gaussian fluctuations can be written \cite{bvm92}: $  \frac{dn(m,z)}{dm} = -\frac{\rho_0}{m^2}\frac{\rmd \ln \nu}{\rmd \ln m}\nu \mathcal{ F}(\nu) $
where $\rho_0$ is the mean matter density today, and $\nu = \delta_{c}(z)/\sigma(m)$ is the normalized amplitude of fluctuations and $\sigma^2(m)$ is the variance of the linear density perturbations within a sphere of comoving radius $R$ that contains mass $m=4\pi\rho_0 R^3/3$. 
\begin{equation}\label{eq:s2m}
  \sigma^2(m,z)=\sigma^2(R,z)=\frac{1}{2\pi^2}\int_0^\infty k^2 P(k,z) W^2(kR) dk
\end{equation}
where $W(kR)$ is the Fourier transform of the top-hat window function, $\delta_c$ represents the critical value of the initial overdensity that is required for collapse at $z$ 
and $P(k,z)$ is the linear power spectrum: $P(k,z)=T(k,z)^2 . D(z)^2 .P_0(k)$
with $P_0(k)$ being the primordial spectrum, $T(k,z)$ a transfer function and $D(z)$ the growth function of cosmological parameters.\\
Another alternative to relate cluster counts to the cosmological  model, also used in this work, is where the root mean square of mass fluctuations is approximated with a power law. 
\begin{equation}
\label{eq:sigm}
\sigma(m,z) =D(z) \sigma_8 \left( \frac{m_8}{m} \right)^{(\frac{n+3}{6})}
\end{equation}
Where D(z) is the growth normalized to unity today, \(m_8\) is the mass contained inside a sphere of \(8h^{-1}~Mpc\) at the mean Universe density and \(n=-2\) is well adapted for cluster scale.
 $\mathcal{ F}(\nu)$ is a function first addressed by Press \& Schechter 1974 (PS) \cite{PS74}
and refined later by many including the one used in this work by Despali et al. 2016 \cite{DP16} (DP):
\begin{equation}
    \nu \mathcal{ F}_{\rm DP}(\nu) =A\sqrt{\frac{2a}{\pi}}\left[1+\left(\frac{1}{a\nu^2}\right)^p\right]\ \exp\left[-\frac{a\nu^2}{2}\right],
  \end{equation}
  where $A=0.3295 $, $a=0.7689$ \& $p=0.2536$ for DP and $A=0.5 $, $a=1$ \& $p=0$ simplifies to PS.
\begin{figure}
 \centering
  \label{fig:sigmamfnl}
\includegraphics[width=0.5\linewidth]{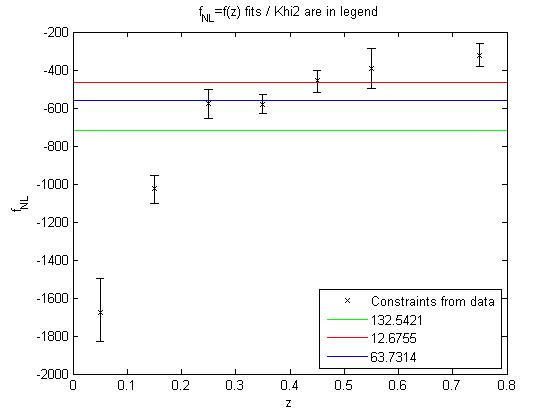}
\caption{ \footnotesize{Showing $f_{NL}$ values (black crosses) that fixes Planck 2013 cluster counts discrepancy for each bin. Showing $f_{NL}$ fitted values for all bins (green line), excluding first bin (blue line) and excluding first two bins (red line). Mass function was calculated using $\sigma(m)$ from Equ.~\ref{eq:sigm}}}
 \end{figure}
 To get the Non-Gaussianity expression of the halo mass function, we follow Loverde et al. 2008 \cite{LoV08} :
   \begin{equation}
\label{equ:massfuncfnl}
\frac{dn_{NG}}{dM}\left(z,M,f_{NL}\right)=\frac{dn_{DP}}{dM}\frac{dn_{PS}/dM(z,M,f_{NL})}{dn_{PS}/dM(z,M,f_{NL}=0)}\,.
\end{equation}
   with
  \begin{eqnarray}
\label{ng_mass_function}
\frac{dn}{dM}\left(M,z\right)=-\sqrt{\frac{2}{\pi}}\frac{\overline{\rho}}{M^{2}}\mathit{e}^{-\delta_{c}^{2}\left(z\right)/2\sigma_{M}^{2}}\left[\frac{d\ln \sigma_{M}}{d\ln M}
\left(\frac{\delta_{c}\left(z\right)}{\sigma_{M}} + \frac{S_{3M}\sigma_{M}}{6} \right. \right. \nonumber \\ 
\left. \left. \times \left(\frac{\delta_{c}^{4}\left(z\right)}{\sigma_{M}^{4}} -2\frac{\delta_{c}^{2}\left(z\right)}{\sigma_{M}^{2}}-1 
 \right) \right) +\frac{1}{6}\frac{dS_{3M}}{d\ln M}\sigma_{M}\left(\frac{\delta_{c}^{2}\left(z\right)}{\sigma_{M}^{2}}-1\right)\right]\,,
\end{eqnarray}
where $S_{3M}=\langle\delta_{M}^{3}\rangle/\langle\delta_{M}^{2}\rangle^{2}\propto f_{NL}$ is the skewness of the smoothed density field we calculate from the fitting formula \cite{ChS10}:
\begin{equation}
S_3 \approx \frac{3.15 \times 10^{-4} \times f_{NL}}{\sigma^{0.838}_M} \label{eq11} 
\end{equation}

\section{Results}

\begin{figure}
\begin{minipage}{0.5\linewidth}
\centerline{\includegraphics[width=0.9\linewidth]{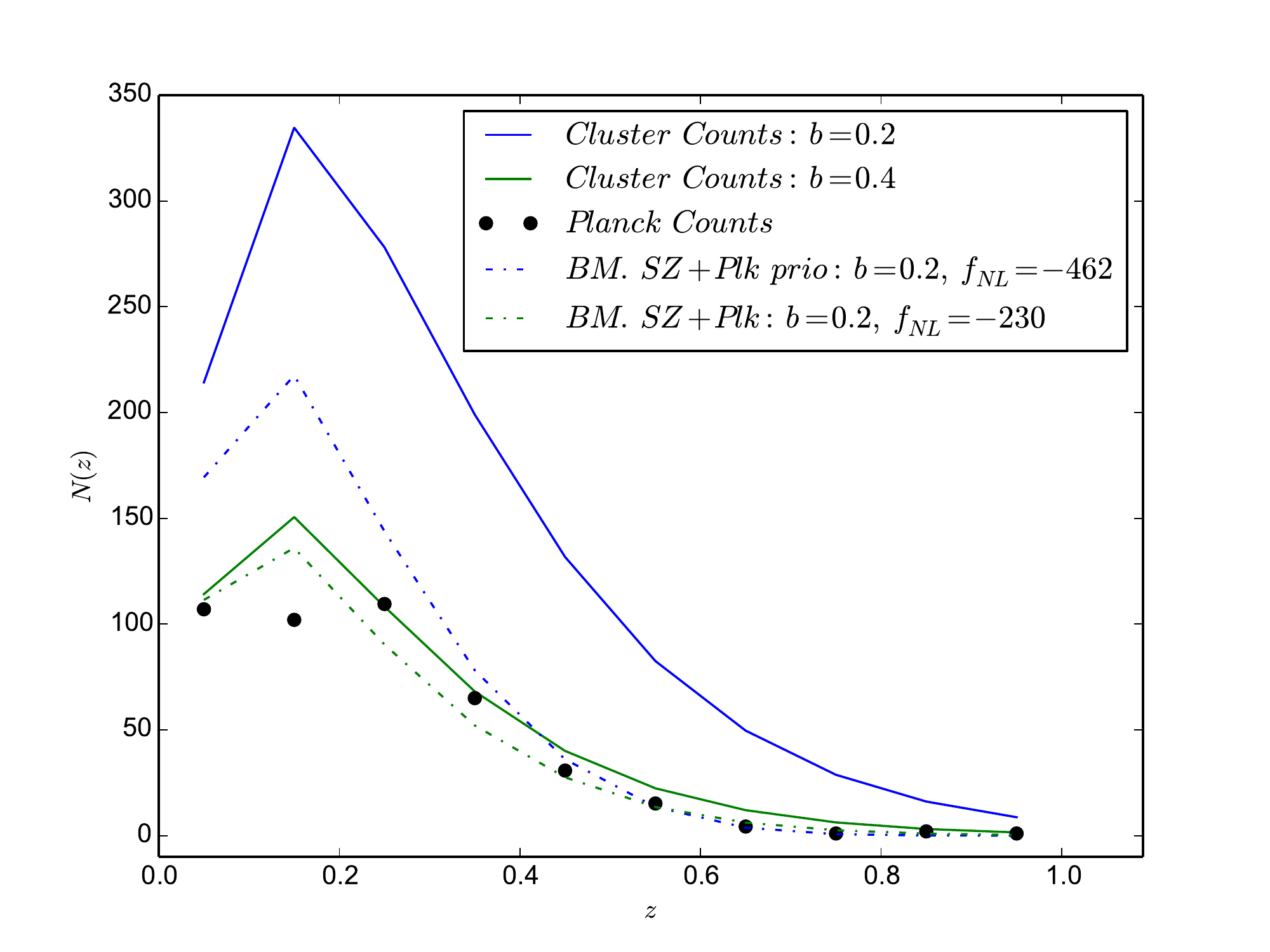}}
\end{minipage}
\hfill
\begin{minipage}{0.5\linewidth}
\centerline{\includegraphics[width=0.9\linewidth]{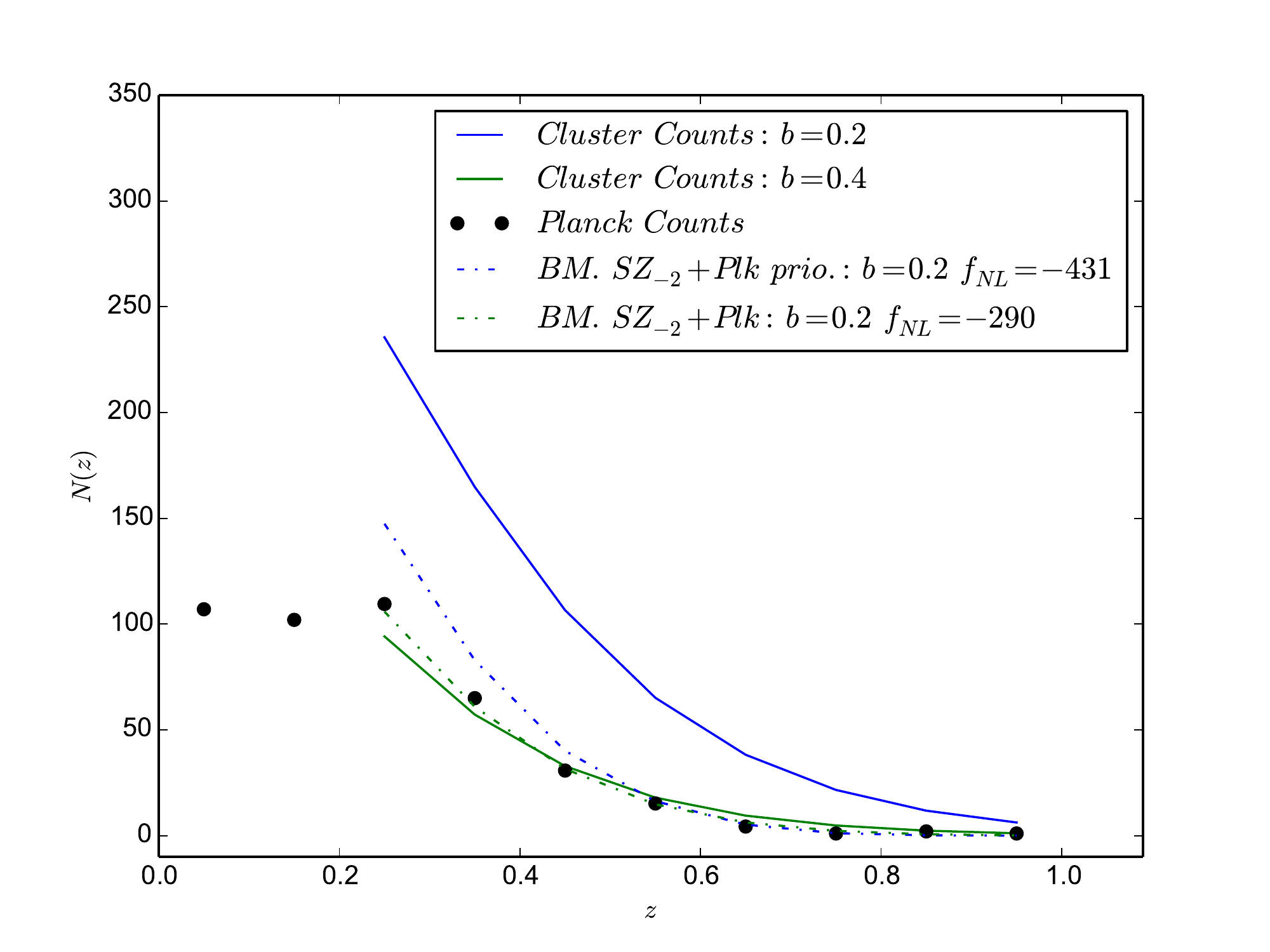}}
\end{minipage}
\hfill
\caption{\footnotesize{Left panel: showing cluster counts from Planck 2015 (black dots), or calculated using SZ Planck calibration (blue line) or calibrated to CMB (green line), from a best fit from a MCMC using SZ clusters sample and Planck priors letting $f_{NL}$ free to vary (dashdot blue line) or a best fit using SZ clusters sample combined with Planck letting cosmological parameters and $f_{NL}$ free to vary (dashdot green line). Right panel: repeating the same analysis but excluding the first two redshift bins. Mass function was calculated using $\sigma(m)$ from Equ.~\ref{eq:s2m}}}
\label{fig:cmbSZfnl}
\end{figure}

To determine what value of $f_{NL}$ could fix the discrepancy we begin by a simple approach comparing the ratio between a Gaussian and Non-Gaussian mass function 
\begin{eqnarray}
theoretical\, ratio&
&=\frac{N^{NG}_{PS}(>m,z,f_{NL})}{N^{NG}_{PS}(>m,z,f_{NL}=0)}
\end{eqnarray}
using the value for $\sigma(m)$ as in Equ.~\ref{eq:sigm}.
The previous ratio is then compared to the ratio of the predicted cluster counts calibrated on CMB power spectrum using SZ mass observables with the cluster mean counts data and errors for each redshift bin.
\begin{eqnarray}
data\, ratio&
&=\frac{data~inferior/mean/superior}{best~model~from~Planck~CMB}
\end{eqnarray}
which yields the $f_{NL}$ values that fix the discrepancy in each bin as we observe in Fig.~1 using Planck 2013 SZ clusters data \cite{Plk14}.\\
Then we fit for $f_{NL}$ in three cases : considering all bins or subtracting the first bin or the first two bins. The last two cases are considered because the first two bins counts do not follow the shape expected by the theortical cluster counts (cf. Fig.~\ref{fig:sigmamfnl}).\\
The three values we obtain : $f_{NL}= -735 \pm{91}$ for all bins, $f_{NL}= -581 \pm{75}$ excluding first bin and $f_{NL}= -485\pm{73}$ excluding first two bins are all ruled out by CMB temperature and polarization angular power spectrum \cite{Plkfnl15} which constrains values for $f_{NL}$ between unities and $\sim20$.\\
We then consider a second case in which $\sigma(m)$ is determined from Equ.~\ref{eq:s2m} 
 entering a cluster counts, function of Non-Gaussianity parameter $f_{NL}$, that is calculated based on the mass function from Equ.~\ref{equ:massfuncfnl}. 
 In this second case we run MCMC chains in order to constrain the Non-Gaussianity parameter $f_{NL}$ from a combination of CMB datasets with fixed best fit cosmological parameters from Planck 2015 mission release \cite{Plk16} and SZ clusters sample datasets with the same cosmological parameters and a calibration factor of value $(1-b)=0.8$. While if this calibration factor is left free and allowed to be constrained by the CMB datasets, it yields a value of $(1-b)=0.6$ so that the discrepancy on $\sigma_8$ could be translated into one on $(1-b)$. Thus when we combine CMB and clusters datasets, keepings for each probe its own calibration, $f_{NL}$ should vary from fiducial null value in order to accommodate the two calibration values.\\
 This is indeed what we observe in left panel of Fig.~\ref{fig:cmbSZfnl} where a $f_{NL}= -462$ (dash dot blue line) will reduce the gap between the SZ calibration cluster counts (blue line) and CMB calibrated cluster counts (green line). Note that higher values like those found in the previous case could reduce more, however they yield unphysical negative counts for high redshift bins. 
To allow more freedom in reducing the discrepancy, we allow the cosmological parameters to vary and combine with CMB datasets. We observe that we reach a lower value for $f_{NL}= -230$ that reduces more the discrepancy even if still ruled out by CMB alone priors on $f_{NL}$. If we stay in the same case but we exclude the first two bins (right panel of Fig.~\ref{fig:cmbSZfnl}), we observe that we reach a lower value for $f_{NL}= -413$ (dash dot blue line) and a higher reduction of the discrepancy with respect to the all bins case. This is not what we observe when we allow the cosmological parameters to vary, where $f_{NL}$ excluding first two bins was found a little bit higher than the all bins case, because the gain of reducing the discrepancy on the second bin is higher than the one we get on the other redshift bins. However, $f_{NL}$ values are still in all cases outside the priors from CMB datasets alone.

\section{Conclusions}

In this work we tested if primordial Non-Gaussianity described by $f_{NL}$ could help fix a discrepancy found on $\sigma_8$ from CMB vs Clusters probe. Following two approaches to fit the best value $f_{NL}$ that could alleviate the tension, we found values of the later that could only reduce the discrepancy but are ruled out by constraints from CMB data. However, Planck mission constraints on non-Gaussian signal were estimated for all scales and were not restricted to sub-intervals on scales which can be associated with the growth of galaxy clusters  \cite{Plkfnl15}, therefore it remains possible that on these scales, the primordial perturbations were non-Gaussian to some extent.

\end{document}